\documentclass{article}
\usepackage{jheppub}
\usepackage[dvipsnames]{xcolor}
\usepackage{mathtools, braket, parskip, slashed,bbm}
\usepackage[none]{hyphenat}
\usepackage{enumitem}
\usepackage{tensor}
\usepackage{bbm}
\usepackage{dsfont}
\usepackage{mathrsfs}
\newif\ifdraft
\draftfalse 
\ifdraft
  \PassOptionsToPackage{draft}{graphicx} 
  \hypersetup{hidelinks} 
\else
	\hypersetup{
		colorlinks = true,
		linkcolor = Mahogany,
		anchorcolor = Mahogany,
		citecolor = Mahogany,
		filecolor = Mahogany,
		urlcolor = Mahogany
	}
  \usepackage{tikz, tikz-feynman}
  \usetikzlibrary{shapes,arrows,positioning,automata,backgrounds,calc,er,patterns}
  \tikzfeynmanset{compat=1.1.0}
    \tikzset{graviton/.style={decorate, decoration={snake, amplitude=.4mm, segment length=1.5mm, pre length=.5mm, post length=.5mm}, double}}
     
\fi
\makeatletter
\newlist{todolist}{itemize}{2}
\setlist[todolist]{label=$\square$}

\graphicspath{{Pictures/}} 

\newcommand{\cl}[1]{\mathcal{#1}}

\renewcommand{\d}{\mathrm{d}}
\newcommand{\dd}{\hat{\mathrm{d}}}
\newcommand{\del}{\hat{\delta}}

\renewcommand{\Re}{\operatorname{Re}}

\newcommand{\nr}{\text{n}}

\def\prd{\ref@{Phys.~Rev.~D}}        

\usepackage{tcolorbox}
\definecolor{airforceblue}{rgb}{0.36, 0.54, 0.66}
\definecolor{azure}{rgb}{0.0, 0.5, 1.0}
\newtcolorbox{tdbox}{colback=airforceblue!40!white,colframe=azure!90!black} 
\newcommand{\td}[1]{
	\if\notesOn1
	\begin{tdbox}
		#1
	\end{tdbox}
	\fi
}
\newcommand{\com}[1]{
	\if\commentsOn1
	\begin{tdbox}
		#1
	\end{tdbox}
	\fi
}

\def\notesOn{1}
\def\commentsOn{1}
\renewcommand{\[}{\begin{equation}\begin{aligned}}
	\renewcommand{\]}{\end{aligned}\end{equation}}

\subheader{\hfill \begin{tabular}{r} \texttt{QMUL-PH-26-12} \end{tabular}}

\title{Memory effect from the scattering of Taub-NUT black holes}

\author {George Doran,}
\author {Ricardo Monteiro, and}
\author {Nathan Moynihan}
\affiliation{Centre for Theoretical Physics, Department of Physics and Astronomy, \\
Queen Mary University of London, E1 4NS, United Kingdom}
\emailAdd{g.e.b.doran@qmul.ac.uk, ricardo.monteiro@qmul.ac.uk, n.moynihan@qmul.ac.uk}

\abstract{
Taub-NUT black holes are somewhat exotic solutions to the vacuum Einstein equations, which have received limited attention in gravitational phenomenology. We use the soft behaviour of scattering amplitudes to compute the memory effect of the waveform resulting from the scattering of Kerr-Taub-NUT black holes. Due to the non-linear nature of gravity, NUT charges introduce intriguing features in the soft dynamics, which have no counterpart in the closely related setting of monopole charges in electromagnetism. In addition to this potentially realistic problem, we also comment on the purely academic problem in complexified gravity of the scattering of self-dual Taub-NUT black holes, which have been discussed recently in the context of celestial holography.
}
\begin{document}

\maketitle

\section{Introduction}
\label{sec:intro}

The direct detection of gravitational waves \cite{LIGOScientific:2016aoc} opened up a new era for probing the nature of gravity, and upcoming observatories will dramatically expand the current capabilities. For instance, the Laser Interferometer Space Antenna (LISA) is expected to allow for the measurement of gravitational memory, a permanent effect caused by the passage of gravitational radiation \cite{Zeldovich:1974gvh,Braginsky:1985vlg,Braginsky:1987kwo,Christodoulou:1991cr,Wiseman:1991ss,Thorne:1992sdb}; see e.g.~\cite{Grant:2022bla,Ghosh:2023rbe,Inchauspe:2024ibs}.

Among the signals from `conventional' objects, there may hide signals from more exotic ones, such as black holes carrying additional charges beyond mass and spin. One such possibility --- and perhaps the earliest historically --- encountered by solving the Einstein equations is that of a NUT charge \cite{Taub:1950ez,Newman:1963yy,1966BAPSS..14..653D}. It would certainly be a surprise to find evidence for (Kerr-)Taub-NUT black holes, in view of their puzzling properties; see \cite{Griffiths:2009dfa} for an overview. In the non-spinning case, the metric can be written as 
\[
\label{eq:TNmetric}
ds^2 = -f(r)(dt + 2n\cos\theta \, d\phi)^2 + \frac{dr^2}{f(r)} + (r^2 + n^2)(d\theta^2 + \sin^2\theta \, d\phi^2)\,,
\]
where
\[
f(r) = \frac{r^2 - 2mr - n^2}{r^2 + n^2}\,.
\]
It reduces to the Schwarzschild solution when the NUT parameter $n$ vanishes. The fact that
\[
g_{t\phi} \approx -2n\cos\theta
\]
at large $r$ has long motivated the interpretation of $n$ as a `gravitomagnetic monopole charge' \cite{Dowker:1974znr}. In fact, this analogy with electromagnetism can be turned into an exact map, where the Taub-NUT solution is a `double copy' of an electromagnetic dyon \cite{Luna:2015paa,Alfonsi:2020lub}. Similarly to the Dirac string of a dyon, the Taub-NUT solution presents a string-like singularity, known as the Misner string, which in the coordinates above lies along $\theta=0,\pi$. For the Dirac string, one may remove the singular behaviour of the gauge potential by an appropriate gauge choice in a given patch. For the Misner string, however, the analogous procedure requires periodicity of the time coordinate, $t\sim t+8\pi n$, which precludes the astrophysical relevance of the solution \cite{Misner:1963fr}. The alternative, where we accept the Misner string (see e.g.~\cite{Bonnor:1969ala,Manko:2005nm}), has stark consequences: there exist closed timelike curves in a region close to the string. This is clear from the metric \eqref{eq:TNmetric}, e.g.~by restricting to trajectories with $t$, $r$ and $\theta$ constant. One may be tempted to summarily discard as unphysical any solution to the Einstein equations possessing closed timelike curves. By doing so, however, we may be missing an important physical feature, or at least a useful counterpoint to conventional gravity. The Misner string is not associated to a singularity of curvature invariants. Moreover, the closed timelike curves are not geodesic \cite{Clement:2015cxa}. The thermodynamics of Lorentzian Taub-NUT black holes has also been discussed; see e.g.~\cite{Hennigar:2019ive,Durka:2019ajz,Bordo:2019tyh,BallonBordo:2019vrn}. In fact, potential observational features of NUT-charged black holes have long been investigated, by studying geodesics on the (Kerr-)Taub-NUT background \cite{Lynden-Bell:1996dpw,Nouri-Zonoz:1997mnd,Wei:2011nj,Rahvar:2002es,Rahvar:2003fh,Bini:2003bm,Shen:2002clt,Liu:2010ja,Pradhan:2014zia,Long:2018tij,Kagramanova:2010bk,Chakraborty:2017nfu,Yang:2020iat}. And yet, advances are well behind the state-of-the-art for conventional black holes, especially when it comes to gravitational wave templates. As we will see, this is not (only) due to lack of interest, but also due to difficulties in understanding the dynamical coupling of a NUT-charged object.

In this paper, we will construct an expression for the gravitational memory resulting from the scattering of two Kerr-Taub-NUT black holes. Our starting point is the new set of tools for perturbative (post-Minkowskian) gravity that directly connect observables such as the scattering angle and the waveform to scattering amplitudes and related objects, e.g.~\cite{Damour:2017zjx,Bjerrum-Bohr:2018xdl,Cheung:2018wkq,Kosower:2018adc,Bern:2019nnu,Kalin:2019rwq,DiVecchia:2021ndb,Herrmann:2021tct,Bern:2020buy,
Bern:2021dqo,Bjerrum-Bohr:2021din,Brandhuber:2021eyq,Dlapa:2021npj,Dlapa:2021vgp,Dlapa:2022lmu,Jakobsen:2023hig,Damgaard:2023ttc,Cristofoli:2021vyo,Jakobsen:2021smu,Brandhuber:2023hhy,Herderschee:2023fxh,Georgoudis:2023lgf,Caron-Huot:2023vxl,Bini:2024rsy,Bohnenblust:2025gir,Bohnenblust:2023qmy,Bern:2025wyd,Driesse:2026qiz}; see \cite{Bern:2022wqg,Kosower:2022yvp,Bjerrum-Bohr:2022blt,Buonanno:2022pgc,DiVecchia:2023frv} for reviews. On-shell tools have already proven their value when considering `dyonic' particles, whether in electromagnetism or in gravity \cite{Huang:2019cja,Emond:2020lwi,Moynihan:2020gxj,Csaki:2020inw,Kim:2020cvf,Monteiro:2020plf,Crawley:2021auj,Monteiro:2021ztt,Caron-Huot:2018ape}. This may seem perplexing, because the notion of S-matrix (and even of asymptotic states) is poorly understood for dyons, where topological features play an important role; see e.g.~\cite{Shnir:2005vvi} for an overview and \cite{Terning:2018udc,Caron-Huot:2018ape,Csaki:2020inw} for more recent discussions. We will not dwell on these subtleties, but will mention why we expect them to be circumvented in our approach.

While our focus is on gravity, it will be useful to compare to the electromagnetic case. Some difficulties, however, are specific to gravity. Notably, the dyonic scattering amplitudes that are strictly needed for the waveform computation are, in the case of electromagnetism, determined by the U(1) electric-magnetic duality. Given the present ignorance of the `nutty' Compton amplitude in gravity, which is a necessary ingredient, we will focus on its soft limit, corresponding to the small frequency part of the waveform. In ordinary electromagnetism and gravity, the leading soft behaviour of the amplitudes has long been determined \cite{Weinberg:1964ew,Weinberg:1965nx}, and its connection to the memory is nowadays well understood \cite{Strominger:2014pwa,He:2014cra,Campiglia:2014yka,Pasterski:2015zua,Strominger:2017zoo}. The dyonic correction to the electromagnetic soft theorem is known \cite{Strominger:2015bla}, and it straightforwardly follows from the U(1) duality. The nutty correction to the soft graviton theorem is more intriguing. Linearised gravity does obey an analogous U(1) duality \cite{Hull:2001iu,Henneaux:2004jw}, but this symmetry is generically broken by non-linear interactions; see e.g.~\cite{Monteiro:2023dev}. And yet, we will argue that the duality is preserved in our soft limit setting and, therefore, in the memory effect.

Finally, motivated by developments in celestial holography, e.g.~\cite{Crawley:2021auj}, we will also discuss briefly the notion of scattering in self-dual electromagnetism and self-dual gravity.

This paper is organised as follows. In section~\ref{sec:KMOCrev}, we will review the on-shell approach to the scattering of dyons. In section~\ref{sec:soft}, we will discuss the difficulty in understanding the dynamical coupling of `nutty' matter, and will propose a workaround in the soft limit. This allows for the computation of the memory effect, which is described in section~\ref{sec:memory}. Changing tack from gravitational phenomenology to the study of self-dual gravity as a toy model, we discuss the scattering of self-dual dyons and black holes in section~\ref{sec:sd}. Finally, we conclude in section~\ref{sec:conclusion}.

\section{Review: dyons in the KMOC formalism}
\label{sec:KMOCrev}

\subsection{Standard KMOC}

Our starting point is the KMOC formalism \cite{Kosower:2018adc,Cristofoli:2021vyo}, a framework for extracting classical observables from scattering amplitudes. We will first consider its standard application, i.e.~{\it without} dyonic charges. Suppose that we can define: (i) the S-matrix, $\mathbb{S}=\mathbbm{1}+iT$, or at least its `classical part'; and (ii) an initial scattering state $|\psi\rangle$. Then, the change in a classical observable $\mathcal{O}$ during a scattering process is
\begin{align}
\Delta \mathcal{O} &= \lim_{\hbar\rightarrow 0}\left[\braket{\psi|\mathbb{S}^\dagger\mathbb{O}\mathbb{S}|\psi} - \braket{\psi|\mathbb{O}|\psi}\right] \nonumber\\
&= \lim_{\hbar\rightarrow 0}\left[i\braket{\psi|[\mathbb{O},T]|\psi} + \braket{\psi|T^\dagger[\mathbb{O},T]|\psi}\right] ,
\end{align}
where the unitarity condition $\mathbb{S}^\dagger\mathbb{S}=\mathds{1}$ was used. The term $\cl{O}(T^\dagger T)$ can be neglected when one works at leading perturbative order, as we will do. 

We are interested in the scattering of two particles with no incoming radiation. The initial state $|\psi\rangle$ involves wavepackets that are sharply peaked around classical momenta $p_1$ and $p_2$, and depends also on the spacelike vector $b^\mu$ encoding the impact parameter. With this starting point, and following the KMOC procedure \cite{Kosower:2018adc}, the leading-order impulse on particle 1 is determined to be\footnote{We denote
$\del(X) \equiv 2\pi \delta(X)$, $\displaystyle\dd^n p \equiv \frac{\d^np}{(2\pi)^n}$.}
\[
\Delta p_1^\mu 
= i\lim_{\hbar\rightarrow 0} \int \hat{d}^4q\,\hat{\delta}(2p_1\cdot q)\,\hat{\delta}(2p_2\cdot q)\,e^{-\frac{i}{\hbar}b\cdot q}\,q^\mu\,\cl{A}_4[p_1,p_2\rightarrow p_1+q,p_2-q],\label{impulse}
\]
where $\cl{A}_4$ is the tree-level $2\mapsto 2$ amplitude.\footnote{Regarding the limit $\hbar\rightarrow0$, the momentum $q$ of the messenger particle (photon or graviton) is of order $\hbar$. The rescaling $q=\hbar\bar q$, where $\bar q$ is a wave-vector, is often used in the literature. Our leading-order case is very simple, so we will abuse notation in later sections, effectively setting $\hbar=1$ after the limit has been taken.} An important feature of this Fourier integral is that, because we are not interested in vanishing impact parameter (which is excluded from our perturbative approach), the final result comes only from the non-analytic piece of $\cl{A}_4$.\footnote{Consider the integral $\int \hat{d}^4q\,\hat{\delta}(2p_1\cdot q)\,\hat{\delta}(2p_2\cdot q)\,e^{\frac{i}{\hbar}b\cdot q}\,f(q)$. For constant $f(q)$, the result is $\propto\delta^2(b)$; note that $b$ is defined such that $b\cdot p_i=0$. Any analytic $f(q)$ then leads to contributions $\partial_{b}\cdots \partial_{b} \delta^2(b)$, schematically.} So $q$ is effectively on-shell, and the residue of $1/q^2$ is determined by unitarity, such that we can make the replacement
\[
\label{eq:A4sub}
\cl{A}_4[p_1,p_2\rightarrow p_1+q,p_2-q] \,\rightsquigarrow\, \frac1{q^2}\sum_\eta\cl{A}_3[p_1;q^{\eta}] \,\cl{A}_3[p_2;(-q)^{-\eta}].
\]
The sum is over the helicities $\eta=\pm$ of the messenger particle (photon or graviton). This is represented in the diagram below, where the dashed line means that $q$ is effectively on-shell, and the grey blobs denote 3-point scattering amplitudes. 
\[
\label{eq:4ptdiag}
\begin{gathered}
\begin{tikzpicture}
  \tikzset{
    Hcircle/.style={
      draw,
      circle,
      fill=gray,
      minimum size=6mm,
      text centered,
    },
  }
  \begin{feynman}
    \vertex (i1) at (-2,  1) {\(p_1\)};
    \vertex (f1) at ( 2,  1) {\(p_1'\)};
    \vertex (i2) at (-2, -1) {\(p_2\)};
    \vertex (f2) at ( 2, -1) {\(p_2'\)};

    \vertex [Hcircle] (vs)     at (0, -1) {};
    \vertex [Hcircle] (vsuppr) at (0,  1) {};
    \vertex (c3) at (-1, 0.33) {};
    \vertex (c4) at ( 1, 0.33) {};

    \diagram*{
      (i1) -- [fermion] (vsuppr) -- [fermion] (f1),
      (i2) -- [fermion] (vs)     -- [fermion] (f2),

      (vsuppr) -- [graviton, edge label=$q$] (vs),

      (c3) -- [dashed, red] (c4)
    };
  \end{feynman}
\end{tikzpicture}
\end{gathered}
\]

The waveform is another observable that can be computed using the KMOC formalism  \cite{Cristofoli:2021vyo}. At large distance (implemented via a stationary phase approximation \cite{Kosower:2022yvp}), the leading-order waveform in electromagnetism is
\begin{align}
F^{\mu \nu}(x) = -\lim_{\hbar\rightarrow 0}\,\frac{\hbar^{3/2}}{2\pi r}\,\Re \sum_\eta \int_0^\infty \dd\omega \, (i\omega) \int &\Big( \prod_{i=1,2} \dd^4 q_i \,\hat{\delta}\left(2p_i \cdot q_i\right) \Big)\,\hat{\delta}^4\left(q_1+q_2+k\right)~\ell^{[\mu} \epsilon^{(\eta) \nu] *}
\nonumber \\ 
& \times e^{-i \omega u}  e^{-\frac{i}{\hbar} b \cdot q_1} \cl{A}_5[p_1, p_2 \rightarrow p_1+q_1, p_2+q_2,\hbar \omega\ell^\eta].
\label{eq:Fwave}
\end{align}
Here, the parametrisation $x^\mu = ut^\mu + r \ell^\mu$ is used, where $t^\mu = (1,\vec{0})$, and $\ell^\mu = (1, \hat{x})$ is the null vector giving the observer's direction; notice that $u=t-r$ is the usual retarded time. There is a sum over the polarisations of the emitted photon, with momentum $k=\hbar \omega\ell$. The expression for gravity is analogous: with $\kappa=\sqrt{32\pi G/\hbar}$,
\begin{align}
R^{\mu \nu\lambda\rho}(x) = \lim_{\hbar\rightarrow 0}\, \frac{\kappa\, \hbar^{1/2}}{4\pi r}\,\Re \sum_\eta \int_0^\infty \dd\omega \,(i\omega)^2 \int &\Big(\prod_{i=1,2} \dd^4 q_i \,\hat{\delta}\left(2p_i \cdot q_i\right) \Big)\,\hat{\delta}^4\left(q_1+q_2+k\right)~\ell^{[\mu} \epsilon^{(\eta) \nu] *}\,\ell^{[\lambda} \epsilon^{(\eta) \rho] *}
\nonumber \\ 
& \times e^{-i \omega u}  e^{-\frac{i}{\hbar} b \cdot q_1} \cl{A}_5[p_1, p_2 \rightarrow p_1+q_1, p_2+q_2, \hbar\omega\ell^\eta].
\label{eq:Rwave}
\end{align}
Similarly to the formula for the impulse, there is in both theories a Fourier integral that effectively means the internal messenger particle is on-shell. Then, we can make the substitution
\begin{align}
\label{eq:A5sub}
\cl{A}_5[p_1,p_2\rightarrow p_1+q_1,p_2+q_2,k^\eta]\,\rightsquigarrow\; & \frac1{q_1^2}\sum_{\eta'}\cl{A}_3[p_1;q_1^{\eta'}] \,\cl{A}_4[p_2;(-q_1)^{-\eta'},k^\eta] \nonumber \\
+ & \, \frac1{q_2^2}\sum_{\eta'}\cl{A}_4[p_1;(-q_2)^{\eta'},k^\eta] \,\cl{A}_3[p_2;q_2^{-\eta'}]
,
\nonumber \\
- & \, \frac1{q_1^2q_2^2}\sum_{\eta',\eta''}\cl{A}_3[p_1;q_1^{\eta'}]\, \cl{A}_3[-q_1^{-\eta'},-q_2^{\eta''},k^\eta] \,\cl{A}_3[p_2;q_2^{-\eta''}]
,
\end{align}
where the 4-point amplitudes are of Compton type.
These three terms are represented diagrammatically as follows.
\[
\label{eq:5ptdiag}
\begin{gathered}
	\begin{tikzpicture}
		\tikzset{
			Hcircle/.style={
				draw,         
				circle,       
				fill=gray,   
				minimum size=4mm,
				text centered,
			},
			}
		\begin{feynman}
		  \vertex (i1) at (-2,  1) {\(p_1\)};
		  \vertex (f1) at ( 2,  1) {\(p_1'\)};
		  \vertex (i2) at (-2, -1) {\(p_2\)};
		  \vertex (f2) at ( 2, -1) {\(p_2'\)};
		
		  \vertex (vtr) at (0.7,  1) {};
		  \vertex [Hcircle] (vs) at (0,  -1) {};  
		  \vertex [Hcircle] (vsuppr) at (0,  1) {};  
		  \vertex (c1) at (0, 0.6) {};
		  \vertex (c2) at (0, 1.4) {};		  
		  \vertex (c3) at (-1, 0.33) {};
		  \vertex (c4) at (1, 0.33) {};
		  \vertex (kend) at (2,  0) {\(k\)};

		  \diagram*{
			(i1) -- [fermion] (vsuppr) -- [fermion] (f1),
		
			(i2) -- [fermion] (vs) -- [fermion] (f2),
		
			(vsuppr) -- [graviton, edge label=$q_1$] (vs) -- [graviton] (kend),

			(c3) -- [dashed,red] (c4)
		  };
		\end{feynman}
		\end{tikzpicture}
	\end{gathered}
	~~~+~
	\begin{gathered}
		\begin{tikzpicture}
			\tikzset{
				Hcircle/.style={
					draw,         
					circle,       
					fill=gray,   
					minimum size=4mm,
					text centered,
				},
				}
			\begin{feynman}
			  \vertex (i1) at (-2,  1) {\(p_1\)};
			  \vertex (f1) at ( 2,  1) {\(p_1'\)};
			  \vertex (i2) at (-2, -1) {\(p_2\)};
			  \vertex (f2) at ( 2, -1) {\(p_2'\)};
			
			  \vertex (vtr) at (0.7,  1) {};
			  \vertex [Hcircle] (vs) at (0,  -1) {};  
			  \vertex [Hcircle] (vsuppr) at (0,  1) {};  
			  \vertex (c1) at (0, 0.6) {};
			  \vertex (c2) at (0, 1.4) {};		  
			  \vertex (c3) at (-1, 0.33) {};
			  \vertex (c4) at (1, 0.33) {}; 
			  \vertex (kend) at (2,  0) {\(k\)};

			  \diagram*{
				(i1) -- [fermion] (vsuppr) -- [fermion] (f1),
			
				(i2) -- [fermion] (vs) -- [fermion] (f2),
			
				(vs) -- [graviton, edge label=$q_2$] (vsuppr) -- [graviton] (kend),
	
				(c3) -- [dashed,red] (c4)
			  };
			\end{feynman}
			\end{tikzpicture}
	\end{gathered}
    ~~~-~
	\begin{gathered}
		\begin{tikzpicture}
			\tikzset{
				Hcircle/.style={
					draw,         
					circle,       
					fill=gray,   
					minimum size=4mm,
					text centered,
				},
				}
			\begin{feynman}
			  \vertex (i1) at (-2,  1) {\(p_1\)};
			  \vertex (f1) at ( 2,  1) {\(p_1'\)};
			  \vertex (i2) at (-2, -1) {\(p_2\)};
			  \vertex (f2) at ( 2, -1) {\(p_2'\)};
			
			  \vertex (vtr) at (0.7,  1) {};
			  \vertex [Hcircle] (vs) at (0,  -1) {};  
			  \vertex [Hcircle] (vsuppr) at (0,  1) {};  
			  \vertex (c1) at (0, 0.6) {};
			  \vertex (c2) at (0, 1.4) {};		  
			  \vertex (c3) at (-0.5, 0.17) {};
			  \vertex (c4) at (0.5, 0.17) {}; 
			  \vertex (c5) at (-0.5, -0.17) {};
			  \vertex (c6) at (0.5, -0.17) {}; 
			  \vertex (kend) at (2,  0) {\(k\)};
            \vertex (k0) at (0,0);

			  \diagram*{
				(i1) -- [fermion] (vsuppr) -- [fermion] (f1),
			
				(i2) -- [fermion] (vs) -- [fermion] (f2),
			
				(vs) -- [graviton, , edge label=$q_2$] (k0), 
                (k0) -- [graviton, , edge label=$q_1$] (vsuppr), 
                (k0) -- [graviton] (kend),
	
				(c3) -- [dashed,red] (c4),
                (c5) -- [dashed,red] (c6)
			  };
			\end{feynman}
			\end{tikzpicture}
	\end{gathered}
\]
The third diagram corrects a double-counting, as each of the first two diagrams admits the singularity corresponding to the third diagram. The point is that $\mathcal A_5$ is given by the right-hand side of \eqref{eq:A5sub} up to terms that are analytical in $q_i$, which give a vanishing contribution to the Fourier integral at non-zero impact parameter.

While we have focused here on the leading order, the impulse and the waveform can in principle be computed at any order if we know the $N$-Compton amplitudes $\cl{A}_{2+N}[p;q_1^{(\eta_1)},\cdots,q_N^{(\eta_N)}]$.

\subsection{Dyonic particles}

Now we consider the scattering of dyonic particles. In electromagnetism, such particles have both electric charge $e_i$ and magnetic charge $g_i$. In gravity, in our context, dyonic means nutty; that is, we consider particles with mass $m_i$ and NUT charge $n_i$.

Naively, the application of the KMOC formalism fails at the first hurdle when applied to dyons already in electromagnetism, because the S-matrix involving more than one dyon species is poorly understood; see e.g.~\cite{Caron-Huot:2018ape}. Even defining the asymptotic states is subtle, due to the property of pairwise helicity \cite{Csaki:2020inw}. Remarkably, however, when we consider the expressions for observables {\it after} the substitutions \eqref{eq:A4sub} and \eqref{eq:A5sub}, corresponding respectively to the diagrams \eqref{eq:4ptdiag} and \eqref{eq:5ptdiag}, these challenges appear to be avoided, because the relevant subamplitudes involve a single dyon species. The validity of this approach is supported by explicit calculations for the leading-order impulse \cite{Huang:2019cja,Emond:2020lwi,Moynihan:2020gxj}, where the result matches that of using the equations of motion. For instance, in gravity, this check is performed by considering a pure-mass probe particle on a Taub-NUT background: the impulse computed via the diagram \eqref{eq:4ptdiag} matches that obtained at leading order from the geodesic equation. In fact, this agreement also holds at higher perturbative orders \cite{Kim:2020cvf,Kol:2021jjc}; an apparent discrepancy in \cite{Kim:2020cvf} has been corrected in work yet to appear.

One may worry that any perturbative approach will conflict with the Dirac-Schwinger-Zwanziger quantisation condition \cite{Dirac:1931kp,Schwinger:1966nj,Zwanziger:1970hk}, or its gravitational counterpart \cite{Bunster:2006rt}. Respectively, the quantised combinations of charges are
\[
e_1g_2-e_2g_1, \qquad u_1\cdot u_2\,(m_1n_2-m_2n_1),
\]
where $u_i^\mu$ is the velocity.\footnote{The gravitational quantisation condition has been derived either in linearised gravity or by considering a pure-mass probe on a NUT-charged background, the latter case leading to a quantisation of $En$, where $E$ is the conserved energy of the probe and $n$ is the NUT charge of the background \cite{Bunster:2006rt,Emond:2021lfy,Kol:2021jjc}. The breaking of U(1) duality by generic gravitational interactions puts a question mark over the quantisation condition's validity more generally.} Our set-up, however, is safe. Firstly, we are interested here in the classical problem, and while we are using a `quantum-first' approach of scattering amplitudes, there is ultimately no quantisation in the classical theory. Secondly, the small dimensionless parameter in our perturbative problem is actually the ratio between the charge combinations above and the angular momentum, as mentioned e.g.~in \cite{Huang:2019cja}, and as nicely exemplified in the explicit results of \cite{Kol:2021jjc}.

To proceed, we need the 3-point and 4-point amplitudes appearing in figure \eqref{eq:5ptdiag}. The 3-point amplitudes in electromagnetism and gravity are given by \cite{Huang:2019cja,Emond:2020lwi}
\[
\label{eq:3ptEM}
\cl{A}_3^\text{EM}[p;q^\eta]  = 2 {\mathcal Q}\, e^{\eta(i\theta+q\cdot a)}\, (\epsilon^{(\eta)}\cdot p), \quad
\text{with}\quad p_\mu=m u_\mu, \;\;{\mathcal Q} = \sqrt{e^2+g^2}, \;\; e^{i\theta}=\frac{e+ i g}{{\mathcal Q}},
\]
and
\[
\label{eq:3ptgrav}
\cl{A}_3^\text{grav}[p;q^\eta]  = \kappa\, e^{\eta(i\theta+q\cdot a)}\, (\epsilon^{(\eta)}\cdot p)^2, \quad
\text{with}\quad p_\mu = {\mathcal M} u_\mu,\;\;
{\mathcal M} = \sqrt{m^2+n^2}, \;\; e^{i\theta} = \frac{m+in}{{\mathcal M}}.
\]
We also included a classical spin parameter $a^\mu$ to allow for rotation \cite{Guevara:2018wpp,Arkani-Hamed:2019ymq,Levi:2015msa,Vines:2017hyw,Chung:2018kqs}.\footnote{Ref.~\cite{Arkani-Hamed:2019ymq} pointed out its relation to the Newman-Janis shift \cite{Newman:1965tw}.} These 3-point amplitudes can be interpreted as an on-shell Fourier transform of their associated stationary classical solution.\footnote{On-shell kinematics at 3 points are not possible with real momenta in Lorentzian signature, but can be defined in a complexified setting, as is commonly used in the amplitudes literature. See \cite{Monteiro:2020plf} for an explicit amplitude/solution map based on real momenta in split signature.} The amplitudes also follow from the U(1) duality of the linearised theories, which acts as
\[
\text{electromagnetism:} \qquad\theta \rightarrow \theta + \vartheta,\qquad
\epsilon^{(\eta)}_\mu \rightarrow  e^{-i\eta\vartheta} \epsilon^{(\eta)}_\mu, \\
\text{gravity:} \qquad\theta \rightarrow \theta + 2\vartheta,\qquad
\epsilon^{(\eta)}_\mu \rightarrow  e^{-i\eta\vartheta} \epsilon^{(\eta)}_\mu.
\]
We note that the momentum $p_\mu = {\mathcal M} u_\mu$ in gravity is duality invariant.

Moving on to the 4-point (Compton) amplitudes, the electromagnetic case is dictated again by the U(1) duality:
\[
\cl{A}_{4,(e,g)}^\text{EM}[p;q^{\eta'},k^\eta] =  e^{i(\eta'+\eta)\theta}\cl{A}_{4,({\mathcal Q},0)}^\text{EM}[p;q^{\eta'},k^\eta].
\]
In fact, this extends straightforwardly to any number of photons, meaning that the impulse and the waveform can be computed at any order in a gauge-invariant manner, with no reference to Dirac strings.

In gravity, however, the U(1) duality is generically broken by the non-linearity \cite{Monteiro:2023dev}. We would like to understand the corresponding Compton amplitude, but for this paper we will focus only on the soft behaviour, which is relevant for the memory effect.

\subsection{Impulse for dyonic particles}

We review here the derivation of the leading-order impulse in electromagnetism and in gravity \cite{Huang:2019cja,Moynihan:2020gxj,Emond:2020lwi,Emond:2021lfy}, both for illustration and because this quantity arises when computing the memory effect. We recall that the impulse is given by \eqref{impulse}, where we make the substitution \eqref{eq:A4sub}, involving the dyonic 3-point and 4-point (Compton) amplitudes given above.

For electromagnetism, and starting with the non-spinning case, the substitution \eqref{eq:A4sub} takes the explicit form (after some algebra)\footnote{We denote $\varepsilon(a,b,c,d)\equiv\varepsilon_{\mu\nu\lambda\rho}a^\mu b^\nu c^\lambda d^\rho$, and $\varepsilon_\mu(b,c,d)\equiv\varepsilon_{\mu\nu\lambda\rho} b^\nu c^\lambda d^\rho$.}
\[
\cl{A}_4^\text{EM}[p_1,p_2\rightarrow p_1+q,p_2-q] \,\rightsquigarrow\, \frac{4}{q^2}\left[(e_1e_2+g_1g_2)(p_1\cdot p_2) + (e_1g_2-e_2g_1)\frac{\varepsilon(p_1,p_2,\nr,q)}{(\nr\cdot q)}\right].
\]
We can, therefore, express the impulse as
\[
	\Delta p_1^\mu = \int\hat{d}^4q\hat{\delta}({u}_1\cdot q)\hat{\delta}({u}_2\cdot q)e^{-iq\cdot b}\,i\,\frac{q^\mu}{q^2} \left((e_1e_2 + g_1g_2)u_1\cdot u_2 - (e_1g_2 - e_2g_1)\frac{\varepsilon(u_1,u_2,\nr,q)}{\nr\cdot q}\right).
\]
These formulas appear to depend on the gauge via the reference vector $\nr_\mu$, which enters via the definition of the polarisation vectors in \eqref{eq:A4sub}, chosen to obey $\epsilon^{(\eta)}\cdot \nr=0$. However, the procedure is really gauge invariant at every step, because we are dealing with scattering amplitudes. Using the Schouten identity, we find
\[
0 &= q^{[\mu}\varepsilon^{\nu\rho\sigma\tau]}{u}_{1\nu}{u}_{2\rho}\nr_\sigma q_\tau = q^{\mu}\varepsilon({u}_1,{u}_2,\nr,q) - (\nr\cdot q)\varepsilon^\mu({u}_1,{u}_2,q) + \cl{O}(q^2, {u}_i\cdot q).
\]
This leads to a formula for the impulse that is manifestly independent of the gauge choice $\nr_\mu$:
\[
	\Delta p_1^\mu = \int\hat{d}^4q\hat{\delta}({u}_1\cdot q)\hat{\delta}({u}_2\cdot q)e^{-iq\cdot b}i\frac{1}{q^2} \left(q^\mu(e_1e_2 + g_1g_2)u_1\cdot u_2 - (e_1g_2 - e_2g_1)\varepsilon^\mu(u_1,u_2,q)\right).
\]
Using the result 
\[
\label{eq:intq}
	i\int\hat{d}^4q\hat{\delta}({u}_1\cdot q)\hat{\delta}({u}_2\cdot q)\frac{q^\mu}{q^2}e^{-iq\cdot b} = -\frac{1}{2\pi\sqrt{\gamma^2-1}}\frac{b^\mu}{b^2}, \qquad \gamma\equiv u_1\cdot u_2,
\]
we obtain
\[
	\Delta p_1^\mu = -\frac{(e_1e_2 + g_1g_2)\gamma}{2\pi\sqrt{\gamma^2-1}}\frac{b^\mu}{b^2} + \frac{(e_1g_2 - e_2g_1)}{2\pi\sqrt{\gamma^2-1}}\frac{\varepsilon^\mu(u_1,u_2,b)}{b^2}.
\]
The charge combinations that appear are manifestly invariant under electric-magnetic duality:
\[
\label{eq:egdual}
e_1e_2 + g_1g_2 = {\mathcal Q}_1 {\mathcal Q}_2 \cos(\theta_1-\theta_2),
\qquad e_1g_2 - e_2g_1= {\mathcal Q}_1 {\mathcal Q}_2 \sin(\theta_1-\theta_2).
\]
This discussion can be extended to dyons with classical spin \cite{Emond:2020lwi}. 

The gravity case follows similarly. Starting from the 3-point amplitudes \eqref{eq:3ptgrav}, and including now also the spin parameters via $a_i^\mu$ for completeness, the result is \cite{Emond:2020lwi}
\begin{align}
\label{eq:impgrav}
	\Delta p_1^\mu = \frac{\kappa^2}{2} {\mathcal M}_1{\mathcal M}_2\; \text{Re}\!\int\hat{d}^4q\, & \hat{\delta}(2{u}_1\cdot q)\hat{\delta}(2{u}_2\cdot q)e^{-iq\cdot b}\frac{1}{q^2} \\ & \times \left(iq^\mu (2\gamma^2-1) - 2 \gamma\, \varepsilon^\mu(u_1,u_2,q)\right)e^{-i(\theta_1-\theta_2)- q\cdot (a_1-a_2)}. \nonumber
\end{align}
The U(1) duality is also manifest in this expression, because $\theta_1-\theta_2$ is invariant. We can perform the integral explicitly using \eqref{eq:intq}, which leads to
\[
\Delta p_1^\mu = -\frac{\kappa^2 {\mathcal M}_1{\mathcal M}_2}{16\pi\sqrt{\gamma^2-1}}\,\mathrm{Re}\left\{e^{-i(\theta_1-\theta_2)}\,\frac{(2\gamma^2-1)\,\tilde b_\perp^\mu + 2i\gamma\,\varepsilon^\mu(u_1,u_2,\tilde b_\perp)}{\tilde b_\perp^{\,2}}\right\},
\]
where $\tilde b_\perp^\mu \equiv b^\mu - i(a_1-a_2)_\perp^\mu$. The projection into the space orthogonal to $u_1$ and $u_2$ is defined as
\[
\label{eq:proj}
a_\perp^\mu \equiv \Pi^{\mu}{}_{\nu}\,a^\nu,\qquad \text{with} \qquad \Pi^{\mu}{}_{\nu} \equiv \delta^{\mu}{}_{\nu} + \frac{u_1^\mu(u_{1\nu} - \gamma u_{2\nu}) + u_2^\mu(u_{2\nu} - \gamma u_{1\nu})}{\gamma^2-1}.
\]

\section{Dyonic soft theorems}
\label{sec:soft}

In this section, we consider the leading soft behaviour of scattering amplitudes in electromagnetism and gravity, extending Weinberg's famous soft factors \cite{Weinberg:1964ew,Weinberg:1965nx} to the case of dyonic/nutty charges. The electromagnetic extension has been addressed in the literature \cite{Strominger:2015bla}, but we will see that in gravity the naive extension conflicts with gauge invariance.

The relevance of the leading soft behaviour of the amplitudes for the purpose of this paper is that it encodes the memory effect \cite{Strominger:2014pwa}. Recalling that the waveform is associated to a 5-point amplitude with an external graviton, as in \eqref{eq:Rwave}, we wish to consider the soft factor $S^{(\eta)}$, such that 
\[
\label{eq:A5SA4}
\cl{A}_5[p_1, p_2 \rightarrow p_1+q_1, p_2+q_2,k^\eta] \simeq S^{(\eta)} \cl{A}_4[p_1, p_2 \rightarrow p_1+q_1, p_2+q_2] \qquad \text{as}\;\; k\rightarrow 0.
\]
This behaviour effectively relates the memory to the impulse, because the latter is determined by the $2\mapsto 2$ amplitude via \eqref{impulse}. In the case of dyonic scattering, the $2\mapsto 2$ amplitude is not fully understood, but as discussed earlier we only need the non-analytic piece.

\subsection{Electromagnetism}

The soft factor in electromagnetism \cite{Weinberg:1965nx,Strominger:2015bla} is
\[
\label{eq:SEM}
S_\text{EM}^{(\eta)}(k)= \sum_i (e_i +i\eta\,g_i)\,\epsilon^{(\eta)}_{\mu}(k)\left(\frac{p_i'^{\mu}}{p_i'\cdot k} - \frac{p_i^\mu}{p_i\cdot k}\right)= \sum_i {\mathcal Q}_ie^{i\eta\theta_i}\,\epsilon^{(\eta)}_{\mu}(k)\left(\frac{p_i'^{\mu}}{p_i'\cdot k} - \frac{p_i^\mu}{p_i\cdot k}\right),
\]
where $p'_i=p_i+q_i$. This expression is manifestly gauge invariant. In fact, gauge invariance requires conservation of both electric and magnetic charges, which is implicit above because we take the dyon with momentum $p_i'$ to have the same charges $e_i$ and $g_i$ as the dyon with momentum $p_i$. For related discussions in the context of asymptotic charges, see \cite{Nande:2017dba,Hosseinzadeh:2018dkh,Choi:2019sjs,Henneaux:2020nxi,McLoughlin:2024ldp}.

One notable feature is that the soft factor is independent of the spin. While the 3-point amplitude \eqref{eq:3ptEM} exhibits a certain similarity between the spin exponential and the dyonic phase, the soft factor is spin independent, because  $e^{\eta k\cdot a}\rightarrow 1$ in the soft limit. The same applies in gravity.



\subsection{Gravity}

The gravitational case is subtler. In the absence of NUT charges ($\theta_i=0$), we must recover Weinberg's soft factor for \eqref{eq:A5SA4},
\[
S_\text{grav}^{(\eta)} \Big|_{\theta_i=0} = \frac{\kappa}{2}\sum_{i} \epsilon^{(\eta)}_{\mu}\epsilon^{(\eta)}_{\nu}\left(\frac{p_i'^{\mu} p_i'^{\nu}}{p_i'\cdot k} - \frac{p_i^\mu p_i^\nu}{p_i\cdot k}\right).
\]
Gauge invariance is easy to check: under $\epsilon^{(\eta)}_{\mu} \mapsto \epsilon^{(\eta)}_{\mu} + \zeta^{(\eta)}k_\mu$, we have
\[
S_\text{grav}^{(\eta)}\Big|_{\theta_i=0} \;\mapsto\; S_\text{grav}^{(\eta)}\Big|_{\theta_i=0} +\kappa\, \zeta^{(\eta)}\epsilon^{(\eta)} \cdot \sum_i (p'_i-p_i).
\]
Conservation of momentum (by which we mean the ``kinematic momentum", associated to ${\mathcal M}_i=\sqrt{m_i^2+n_i^2}$) implies
\[
\sum_i (p'_i-p_i)=\sum_i q_i=0
\]
in the soft limit.\footnote{Recall the momentum conservation condition $q_1+q_2+k=0$ in the waveform expressions \eqref{eq:Fwave} and \eqref{eq:Rwave}.} The naive extension to NUT charges is
\[
\label{eq:Snaive}
S_\text{naive grav}^{(\eta)}= \frac{\kappa}{2}\sum_{i} e^{i\eta\theta_i}\,\epsilon^{(\eta)}_{\mu}\epsilon^{(\eta)}_{\nu}\left(\frac{p_i'^{\mu} p_i'^{\nu}}{p_i'\cdot k} - \frac{p_i^\mu p_i^\nu}{p_i\cdot k}\right),
\]
where $e^{i\theta_i}=(m_i+in_i)/{\mathcal M_i}$. However, gauge invariance now fails:
\[
\sum_i e^{i\eta\theta_i} (p'_i-p_i) \neq 0.
\]
Indeed, nutty phases violate the conventional universality of soft gravitational coupling. They appear to imply that both the ``mass momentum" and the ``NUT momentum" (respectively, the real and imaginary parts of the expression above) must be conserved by the leading-order interaction. However, this contradicts the impulse result, which implies only conservation of the ``kinematic momentum", $\sum_i\Delta p_i^\mu=0$,\footnote{Notice the scattering is elastic at leading order.} as opposed to $\sum_i e^{i\eta\theta_i}\Delta p_i^\mu=0$.

We have already alluded to various sources of subtleties with dyons, and here is one other example that the puzzles are more dramatic in gravity. At this stage, one may be tempted to discard the possibility of dynamical NUT-charged objects, or perhaps to make a draconian restriction of the allowed cases to special kinematical configurations or to objects with identical dyonic phase $\theta_i$. 

The scattering amplitudes approach suggests a natural gauge-invariant correction, however, which like \eqref{eq:Snaive} preserves the U(1) duality in the soft limit. Consider the expression
\[
\label{eq:totSgravb}
S_\text{grav}^{(\eta)} = \frac{\kappa}{2}\sum_i e^{i\eta\theta_i}\,\epsilon^{(\eta)}_{\mu}\epsilon^{(\eta)}_{\nu}\left(\frac{p_i'^{\mu} p_i'^{\nu}}{p_i'\cdot k} - \frac{p_i^\mu p_i^\nu}{p_i\cdot k}- \frac{q_i^\mu q_i^\nu}{q_i\cdot k}\right),
\]
where $q_1=-q_2$ as $k\rightarrow0$. The condition for gauge invariance is now $\sum_i e^{i\eta\theta_i} (p'_i-p_i-q_i)=0$; recall that $p_i'=p_i+q_i$. This expression is motivated by the diagrams in \eqref{eq:5ptdiag}, with each term ($i=1,2$) seemingly associated to one of the first two diagrams (the third diagram does not contribute to the soft factor in the standard setting). But the expression is intriguing, because of the coupling of the soft graviton to the exchanged graviton with momentum $q_i$. In the case $\theta_i=0$, this coupling cancels among the $i=1,2$ contributions, leading to the Weinberg soft factor. The standard notion of soft factorisation is that it involves only the external particles. Moreover, the coupling to the exchanged graviton here comes with unexpected nutty phases, which naively contradict unitarity. And yet, the physics of nutty couplings must necessarily break usual assumptions, and there is no current understanding of unitarity in this context. Gauge invariance motivates the generalisation of the soft factor written above, which is in fact quite simple. 
One can, however, spot an additional puzzling feature, namely the pole at $q\cdot k=0$. We will discuss this point later on.

It would be interesting to apply our argument based on scattering amplitudes to the recent works \cite{Godazgar:2019dkh,Godazgar:2018qpq,Kol:2019nkc,Kol:2020ucd,Oliveri:2020xls}, which have a similar motivation, alongside older works, e.g.~\cite{Ashtekar:1982}. These deal with NUT parameters in the context of asymptotic symmetries, by considering {\it dual} charges at null infinity. We expect our observations on gauge invariance to be relevant also if massless nutty particles are considered.

\section{Memory effect}
\label{sec:memory}

Having discussed the soft behaviour of the amplitudes, we will in this section use it to determine the \textit{memory effect}. This effect is the measurable imprint left over by the passage of radiation on suitable probes, e.g.~a pair of charges in electromagnetism or a pair of freely falling bodies in gravity. It is typically expressed at future null infinity as a function of celestial coordinates $z,\bar{z}$, or equivalently in terms of a null vector $\ell^\mu = (1,\hat{x})$, which specifies the direction to the distant observer measuring the effect.

\subsection{Electromagnetism}

Let us start with electromagnetism. We use the standard parametrisation $x^\mu = u\,t^\mu + r\,\ell^\mu$, and the null tetrad
$\{\ell^\mu,\tilde\ell^\mu,m^\mu,\bar m^\mu\}$ adapted to the observation direction; see appendix~\ref{NPFormalism} for our conventions. The component of the field strength tensor
relevant for a transverse detector on the celestial sphere is obtained by contracting with
$\tilde\ell^\mu$ and the sphere tangent vectors $e_A^\mu=\partial_A \ell^\mu$:
\[
\label{eq:FA}
F_A(u,r,z,\bar z)
\;\equiv\;
F_{\mu\nu}(u,r,z,\bar z)\,\tilde\ell^\mu e_A^\nu,
\]
where we take $F_{\mu\nu}$ to be the soft expansion of the large-$r$ expression \eqref{eq:Fwave}.
The electromagnetic memory is then defined as a one-form $\mathcal{E}_Adx^A$ on the celestial sphere, with components
\[
\mathcal{E}_A(z,\bar z)
\;\equiv\; \lim_{r\rightarrow\infty} 8\pi r
\int_{-\infty}^{+\infty}\!du\;F_A(u,r,z,\bar z).
\]
This is in principle directly measurable as a velocity kick: for a slowly-moving test (electric) charge $q$ of mass $m$ at
large $r$, the transverse velocity kick is \cite{Bieri:2013hqa,Susskind:2015hpa}
\[
\label{eq:vEM}
\Delta v_A = \frac{q}{m}\,\frac{\mathcal{E}_A(z,\bar z)}{8\pi r}.
\]

We will now determine the memory, starting from the waveform formula \eqref{eq:Fwave}. The classical limit is straightforward at tree level, so we will drop $\hbar$ here, and consider only the classical piece of the amplitudes. Using \eqref{eq:FA}, we have
\begin{align}
F_A(u,r,z,\bar z)
&= -\frac{1}{2\pi r}\,\Re\!\int_0^\infty\dd\omega\,\omega\,e^{-i\omega u}
\int d\mu\;e^{-ib\cdot q_1}\,iV_A(\omega;z,\bar z),
\\[4pt]
V_A(\omega;z,\bar z)
&\equiv \sum_{\eta} V_A^{(\eta)}
\;=\;\sum_{\eta}
\Big(\ell^{[\mu}\epsilon^{(\eta)\nu]*}\tilde{\ell}_{\mu}e_{A\nu}\Big)\,
{\mathcal A}_5\big[p_1,p_2\!\rightarrow\!p_1\!+\!q_1,p_2\!+\!q_2,\omega\ell^{\eta}\big],
\label{eq:VA}
\end{align}
where, for brevity, we also denote $\d\mu\equiv \big(\prod_{i=1,2} \dd^4 q_i \,\hat{\delta}\left(2p_i \cdot q_i\right) \big)\hat{\delta}^4\left(q_1+q_2+k\right)$. Only the zero-frequency limit contributes to the memory:
\begin{align}
\mathcal{E}_A(z,\bar z) & = -4 \int_{-\infty}^{+\infty}\!du\;\Re\!\int_0^\infty\dd\omega\,\omega\,e^{-i\omega u}
\int d\mu\;e^{-ib\cdot q_1}\,iV_A(\omega;z,\bar z)
\nonumber \\
& = - 2\int d\mu\;e^{-ib\cdot q_1}\,i \,\lim_{\omega\rightarrow 0} \,\omega \,V_A(\omega;z,\bar z),
\end{align}
Hence, we are interested in the soft limit of the five-point amplitude,
\[
{\mathcal A}_5^{(\eta)} \;\approx\;\epsilon_\alpha^{(\eta)}\,S^{(\eta)\,\alpha}\,{\mathcal A}_4,\qquad 
S^{(\eta)\,\alpha}
=\sum_i Q_i\,e^{i\eta\theta_i}\!\left(\frac{p_i'^{\alpha}}{p_i'\!\cdot k}-\frac{p_i^{\alpha}}{p_i\!\cdot k}\right).
\] 
To proceed, we first project to sphere indices using the relations in appendix~\ref{NPFormalism}, such that
\[
\ell^{[\mu}\epsilon^{(\eta)\nu]*}\tilde{\ell}_{\mu}e_{A\nu}
=\ell^{[\mu}\epsilon^{(-\eta)\nu]}\tilde{\ell}_{\mu}e_{A\nu}
=-m_A^{(-\eta)}.
\]
We define
\[
S^{(\eta)\,B}\equiv e^B{}_\alpha S^{(\eta)\,\alpha},\qquad
p_i^{B}\equiv e^B{}_\mu p_i^\mu,
\]
and the polarisation vectors are written as  $\epsilon_\mu^{(\eta)}=m_A^{(\eta)}e^A{}_\mu$.
Then, the helicity sum in \eqref{eq:VA} collapses to a duality-rotated projector on the celestial sphere:
\begin{align}
V_A
=-\sum_{\eta,i} Q_i\,e^{i\eta\theta_i}\,m_A^{(-\eta)}m_B^{(\eta)} \,S^{(\eta)\,B} {\mathcal A}_4
=-\sum_i Q_i\,P^{(\theta_i)}_{AB} \left(\frac{p_i'^{B}}{p_i'\!\cdot k}-\frac{p_i^{B}}{p_i\!\cdot k}\right)\, {\mathcal A}_4,
\label{eq:dyonSphereProjector}
\end{align}
where we used
\[
P^{(\theta)}_{AB}\equiv \sum_{\eta} e^{i\eta\theta}\,m_A^{(-\eta)}m_B^{(\eta)}
=
e^{i\theta}\bar m_A m_B+e^{-i\theta} m_A\bar m_B
=\cos\theta\,\gamma_{AB}-\sin\theta\,\varepsilon_{AB}.
\]
Equivalently, writing $e_i\equiv Q_i\cos\theta_i$ and $g_i\equiv Q_i\sin\theta_i$, we find
\[
Q_i P^{(\theta_i)}_{AB}X^B = e_i\,X_A - g_i\, \varepsilon_{AB}X^B .
\]

To obtain the memory, we take $k^\mu=\omega\ell^\mu$, and expand $p_i'=p_i+q_i$ with $q_i\ll p_i$, corresponding to the classical regime:
\[
\frac{p_i'^{B}}{p_i'\!\cdot k}-\frac{p_i^{B}}{p_i\!\cdot k}
=\frac{1}{\omega}\left(
\frac{q_i^{B}}{\ell\cdot p_i}
-\frac{(\ell\cdot q_i)\,p_i^{B}}{(\ell\cdot p_i)^2}
\right)+\mathcal{O}(q^2).
\]
The memory is then
\[
\mathcal{E}_A = 2\int d\mu\;e^{-ib\cdot q_1}\,i \sum_i Q_i\,P^{(\theta_i)}_{AB} \left(
\frac{q_i^{B}}{\ell\cdot p_i}
-\frac{(\ell\cdot q_i)\,p_i^{B}}{(\ell\cdot p_i)^2}\right)\, {\mathcal A}_4.
\]
Using the definition of the impulse
\[
i\int d\mu\;e^{-ib\cdot q_1}\,q_i^{B}\,{\mathcal A}_4 \;=\;\Delta p_i^{B},
\]
we obtain
\[
\mathcal{E}_A
=2\sum_i (e_i\,\delta_A{}^{B}-g_i\,\varepsilon_A{}^{B})
\left(
\frac{\Delta p_{iB}}{\ell\cdot p_i}
-\frac{(\ell\cdot\Delta p_i)\,p_{iB}}{(\ell\cdot p_i)^2}
\right).
\]
Any one-form on $S^2$ decomposes into a gradient plus a curl. This allows us to write
\[
\mathcal{E}_A = D_A\Phi-\varepsilon_{AB}D^B\varphi,
\qquad
\Phi=\sum_i 2e_i\,\frac{\ell\cdot\Delta p_i}{\ell\cdot p_i},
\qquad
\varphi=\sum_i 2g_i\,\frac{\ell\cdot\Delta p_i}{\ell\cdot p_i},
\]
where we used that $D_A$ acts as $ e_A^\mu\frac{\partial}{\partial \ell^\mu}$ on scalar functions of $\ell^\mu$, such that
\[
D_A(\ell\cdot p_i)=p_{iA},\qquad
D_A\!\left(\frac{\ell\cdot\Delta p_i}{\ell\cdot p_i}\right)
=\frac{\Delta p_{iA}}{\ell\cdot p_i}-\frac{(\ell\cdot\Delta p_i)\,p_{iA}}{(\ell\cdot p_i)^2}.
\]
An equivalent form is
\[
\mathcal{E}_A
=\Re\sum_i 2Q_i e^{i\theta_i}\,
\big(D_A+i\varepsilon_{AB}D^B\big)\,\frac{\ell\cdot\Delta p_i}{\ell\cdot p_i}.
\]
The duality covariance is manifest. According to the equation \eqref{eq:vEM}, the magnetic charges $g_i$ introduce a curl component to the velocity kick caused by the passage of the radiation.

\subsection{Gravity}
In gravity, the memory effect is the permanent displacement of a pair of particles in free-fall at $\mathscr{I}^+$, caused by the passage of gravitational radiation. The change in displacement is found by integrating the geodesic deviation equation. It is expressed in terms of the memory tensor $\mathcal{E}_{AB}(z,\bar z)$, the large distance $r$ to the scattering event, and the initial separation $\xi^A$ on the celestial sphere as
\[
\Delta \xi_A = \frac{\mathcal{E}_{AB}(z,\bar z)}{4\pi r}\,\xi^B,
\]
where\footnote{This quantity is often denoted by $\Delta C_{AB}$, $\Delta^{GR}_{AB}$ or $\Delta \sigma_{AB}$ in the literature \cite{Madler:2016ggp,Prabhu:2022zcr}.}
\[
\mathcal{E}_{AB}(z,\bar z) \equiv 
-\lim_{r\rightarrow\infty} 4\pi r \int_{-\infty}^{+\infty}du\int_{-\infty}^{u}\!du'~ R_{\mu\nu\rho\sigma}(u',r,z,\bar z)\,
\tilde\ell^\mu e_A^\nu\,\tilde\ell^\rho e_B^\sigma.
\]

To compute $\mathcal{E}_{AB}$ from the five-point waveform formula \eqref{eq:Rwave}, it is useful to introduce the object
\[
w_{AB}\equiv
\tilde\ell^\mu e_A^\nu\,\tilde\ell^\rho e_B^\sigma\sum_{\eta}\,\ell_{[\mu}\epsilon_{(\eta)\nu]}^*\, \ell_{[\rho}\epsilon_{(\eta)\sigma]}^*\,{\mathcal A}_5^{(\eta)},
\]
such that
\[
\mathcal{E}_{AB}  = -\kappa \int_{-\infty}^{+\infty}du\int_{-\infty}^{u}\!du'\, \Re\!\int_0^\infty\dd\omega\,
(i\omega)^2e^{-i\omega u'}\!\int d\mu\;\,e^{-ib\cdot q_1} w_{AB}.
\]
We now use the identity
\[
\int_{-\infty}^{u}\!du'\,e^{-i\omega u'} = e^{-i\omega u}\left(\pi \delta(w)+i\,\text{PV} \frac1{\omega}\right),
\]
and note that the piece with $\delta(w)$ cannot contribute to the memory tensor, because $(i\omega)^2 w_{AB} \propto \omega $ for small $\omega$. This leads to
\begin{align}
\mathcal{E}_{AB} &= \kappa \int_{-\infty}^{+\infty}du \, \Re\!\int_0^\infty \dd\omega\, e^{-i\omega u}\, (i\omega) \int d\mu\;\,e^{-ib\cdot q_1}\,
w_{AB}\nonumber \\
&= \frac{\kappa}{2} \int d\mu\;\,e^{-ib\cdot q_1}\,\,\lim_{\omega\rightarrow 0} \,
(i\omega)\, w_{AB}.
\label{eq:Ew}
\end{align}
As expected, the memory is captured by the soft limit. In this limit, ${\mathcal A}_5^{(\eta)}=S^{(\eta)}{\mathcal A}_4$. Writing $S^{(\eta)}=\frac{\kappa}{2}\,\epsilon_\alpha^{(\eta)}\epsilon_\beta^{(\eta)}S^{(\eta)\alpha\beta}$, and using $\epsilon_{(\eta)\nu}^*=\epsilon_{(-\eta)\nu}$, we have
\[
 w_{AB}
=\frac{\kappa}{2}\, \tilde\ell^\mu e_A^\nu\,\tilde\ell^\rho e_B^\sigma\sum_{\eta}\,\ell_{[\mu}\epsilon_{(-\eta)\nu]}\, \ell_{[\rho}\epsilon_{(-\eta)\sigma]}\,
\epsilon_\alpha^{(\eta)}\epsilon_\beta^{(\eta)}S^{(\eta)\alpha\beta}{\mathcal A}_4.
\]
We will again follow the tetrad and celestial sphere conventions in appendix~\ref{NPFormalism}. We also factor out the nutty phases from the kinematic dependence of the soft factor, defining $S_i^{CD}$ such that $\sum_ie^{i\eta\theta_i}S_i^{CD}= e^C{}_\alpha e^D{}_\beta S^{(\eta)\alpha\beta}$. Finally, we obtain
\[
\label{eq:wABdef}
w_{AB}
=\frac{\kappa}{2}\sum_{\eta,i} e^{i\eta\theta_i}\,
m_A^{(-\eta)}m_B^{(-\eta)}m_C^{(\eta)}m_D^{(\eta)}\,S_i^{CD}{\mathcal A}_4.
\]
We use the dyad decomposition of the sphere metric \eqref{spheremetricdyads} to perform the helicity sum, leading to
\[\label{eq:wABtracefree}
w_{AB}
&=\frac{\kappa}{4}\sum_i\Big[
\cos\theta_i\big(\gamma_{AC}\gamma_{BD}-\varepsilon_{AC}\varepsilon_{BD}\big)
-\sin\theta_i\big(\varepsilon_{AC}\gamma_{BD}+\varepsilon_{BD}\gamma_{AC}\big)
\Big]S_i^{CD}{\mathcal A}_4 \\
&=\frac{\kappa}{4}\sum_i \Big[\cos\theta_i \,E_{ABCD} -\sin\theta_i\,O_{ABCD} \Big] S_i^{CD}{\mathcal A}_4.
\]
Here, we have defined the following tensor structures on the sphere
 \[
 E_{ABCD}\equiv \gamma_{AC}\gamma_{BD}+\gamma_{AD}\gamma_{BC}-\gamma_{AB}\gamma_{CD}, \qquad 
 O_{ABCD} \equiv \varepsilon_{AC}\gamma_{BD}+\varepsilon_{BD}\gamma_{AC},
 \]
 and simplified the result using $\varepsilon_{AC}\varepsilon_{BD}=\gamma_{AB}\gamma_{CD}-\gamma_{AD}\gamma_{BC}$.
We now expand the kinematic pieces of the soft factor \eqref{eq:totSgravb}, namely $S_i^{AB}$, in the classical limit $q_i\ll p_i$:
\[
S_i^{AB} \equiv \frac{p_i'^A p_i'^B}{p_i'\cdot k} - \frac{p_i^A p_i^B}{p_i\cdot k} - \frac{q_i^A q_i^B}{q_i\cdot k}
= \frac{p_i^{(A} q_i^{B)}}{p_i\!\cdot k} - \frac{(k\cdot q_i)\,p_i^{A}p_i^{B}}{(p_i\cdot k)^2} - \frac{q_i^A q_i^B}{q_i\cdot k}+ \mathcal{O}(q^2),
\label{eq:SiABsoft}
\]
which follows from expanding $p_i' = p_i + q_i$ to leading order in $q_i$. Notice that our symmetrisation convention \eqref{eq:symcon} does not include $1/2$.

To recap, we wish to compute the integral \eqref{eq:Ew}, where $w_{AB}$ is given by \eqref{eq:wABtracefree}, with $S_i^{AB}$ approximated as in \eqref{eq:SiABsoft}. We start with the $\int d\mu$ integral. On the right-most side of \eqref{eq:SiABsoft}, we have the terms $\sim q$ and the term $\sim q q/q$. Dealing with the terms $\sim q$ is straightforward, due to the impulse identity,
\[
i\int d\mu\;e^{-ib\cdot q_1}\,q_i^\mu {\mathcal A}_4 = \Delta p_i^\mu.
\]
For the term $\sim q q/q$, we can show that
\[
\label{eq:Ii}
I_i^{\mu\nu}\equiv i\int d\mu\;e^{-ib\cdot q_1}\,\frac{q_i^\mu q_i^\nu}{q_i\cdot k}\, {\mathcal A}_4 = \frac1{k_\perp^2}\left(\Delta p_i^{(\mu} k_\perp^{\nu)}-\eta_\perp^{\mu\nu} \Delta p_i\cdot k\right),
\]
where the subscript $\perp$ denotes projection to the space orthogonal to the $p_i={\mathcal M}_i u_i$, already defined in \eqref{eq:proj}. The inverse metric on this space is $\eta_\perp^{\mu\nu} = \Pi^{\mu}{}_{\alpha} \eta^{\alpha\nu}$, which we can also write as
\[
\eta_\perp^{\mu\nu} = \varepsilon_\perp^{\mu\alpha}\varepsilon_\perp^\nu{}_{\alpha}\,, \qquad \text{where}
\quad \displaystyle \varepsilon^{\mu\nu}_\perp\equiv \varepsilon^{\mu\nu\alpha\beta}\frac{u_{1\alpha}u_{2\beta}}{\sqrt{\gamma^2-1}}\,.
\]

This result for $I_i^{\mu\nu}$ follows from expressing it in a basis $\{\Delta p_i^{(\mu}k_\perp^{\nu)},k_\perp^{(\mu}k_\perp^{\nu)},\eta_{\perp}^{\mu\nu}\}$, where we note that $\Delta p_i^\mu=\Delta p_{i\perp}^\mu$, and then imposing the constraints $k_{\perp\mu}I_i^{\mu\nu}=k_\mu I_i^{\mu\nu}=\Delta p_i^\nu$ and $\eta_{\perp \mu\nu}I_i^{\mu\nu}=\eta_{\mu\nu}I_i^{\mu\nu}=0$. The latter constraint requires further comment. The quantity $\eta_{\mu\nu}I_i^{\mu\nu}$ has distributional support at $\varepsilon_\perp(b,k)=\varepsilon_\perp^{\mu\nu}b_\mu k_\nu=0$; see also appendix~\ref{app:locus}. Hence, the result \eqref{eq:Ii} is not valid on that strict locus. For $|k_\perp|\neq0$ (we comment later on $|k_\perp|=0$), this locus corresponds to a great circle on the celestial sphere.\footnote{Consider a reference frame where the nutty particles move initially along the $z$ direction, and the impact parameter vector lies along the $x$ direction. Then the special locus is such that $k$ vanishes in the $x$ direction.} Let us make two more remarks concerning this locus. Firstly, we did not yet discuss the $i\epsilon$ prescription for the extra terms in the soft factor \eqref{eq:totSgravb}. Notice that
\[
\frac1{k\cdot q_i\pm i0} = \text{PV}\left(\frac1{k\cdot q_i}\right) \mp i\pi\delta(k\cdot q_i).
\]
The result \eqref{eq:Ii} effectively picks up the principal value, while the $\delta(k\cdot q_i)$ leads again to a contribution with $\delta(\varepsilon_\perp(b,k))$. This can be checked by expressing the integral with $\delta(k\cdot q_i)$ in the basis mentioned above. Secondly, we note that we have no hope of fixing the total contribution to $I_i^{\mu\nu}$ at $\varepsilon_\perp(b,k)=0$. Early on, when we discussed the impulse, we argued that we only need the non-analytic part of ${\mathcal A}_4$, but the presence of the factor $\frac1{k\cdot q_i}$ in $I_i^{\mu\nu}$ means that we now also need the analytic part of ${\mathcal A}_4$, which is unknown. Fortunately, the factor $\frac1{k\cdot q_i}$ also means that the contribution from the analytic part of ${\mathcal A}_4$ would be solely supported at $\varepsilon_\perp(b,k)=0$. To summarise, we cannot determine $I_i^{\mu\nu}$ at exactly $\varepsilon_\perp(b,k)=0$. Given that this locus is a zero-measure set on the celestial sphere, we will not discuss it further.

Returning to \eqref{eq:Ew}, we have
\[
{\mathcal E}_{AB} = 
\frac{\kappa^2}{8}\sum_i \Big[\cos\theta_i \,E_{ABCD} -\sin\theta_i\,O_{ABCD} \Big] \mathcal{I}_i^{CD},
\]
with
\[
\mathcal{I}_i^{CD}&\equiv \int d\mu \,(i\omega)\,e^{-ib\cdot q_1}\,S_i^{CD}{\mathcal A}_4 \\
&=  \left[\frac{p_i^{(C}\Delta p_i^{D)}}{p_i\!\cdot \ell}
-\frac{(\ell\cdot \Delta p_i)\,p_i^{C}p_i^{D}}{(p_i\cdot \ell)^2}  -
\frac{\ell_\perp^{(C}\Delta p_i^{D)}}{\ell_\perp^2}
+\frac{\eta_\perp^{CD}\,\ell_\perp\cdot\Delta p_i}{\ell_\perp^2}
\right],
\label{eq:IiAB}
\]
where we have set $k^\mu=\omega\ell^\mu$. We can rewrite the result as
\[
{\mathcal E}_{AB}=  \frac{\kappa^2}{8}\Re\sum_i e^{i\theta_i}\Big[ {E}_{ABCD} +i{O}_{ABCD}  \Big] \mathcal{I}_i^{CD}\, .
\]
It is also possible to write $\mathcal{I}_i^{AB}$ in terms of derivatives on the sphere of potentials $\Phi_i$ as follows:
\begin{equation}
    \mathcal{I}_i^{AB} = D^A D^B \Phi_i \,,
\end{equation}
where 
\[
\Phi_i =
(\ell\cdot \Delta p_i)\,\log\!\left(\frac{p_i\cdot \ell}{|\ell_\perp|}\right) + \varepsilon_\perp(\ell_\perp,\Delta p_i)\arctan\!\left(\frac{\varepsilon_\perp(\ell_\perp,\Delta p_i)}{\ell\cdot \Delta p_i}\right) \,.
\]
The potentials are finite on the celestial sphere (note that $\ell\cdot\Delta p_i=\ell_\perp\cdot\Delta p_i$), but they are not analytic at the two points on the sphere where $|\ell_\perp|=0$. It is clear from \eqref{eq:IiAB} that the memory tensor diverges there. This indicates a breakdown of our soft/perturbative approach in this region, corresponding to the divergence of the terms in the soft factor \eqref{eq:totSgravb} that are demanded by gauge invariance. One reason to expect such a breakdown is the following. Notice that $q:=q_1=-q_2$ in the soft factor \eqref{eq:totSgravb} lives in a two-dimensional space ($q=q_\perp$) enforced by the measure $d\mu$ of the integral we performed in this section. If the observer of the memory has a position such that $|\ell_\perp|\ll1$, then $q\cdot k$ is extra small, i.e.~it is small not just because the frequency is small, but also because of the direction. On the other hand, the terms $\sim1/(q\cdot k)$ in the soft factor are expected to arise from a propagator $\sim 1/(q\pm k)^2=1/(q^2\pm2 k\cdot q)$. We mentioned in the beginning that $q$ is effectively on-shell, which justifies $\sim1/(2q\cdot k)$, but if $q\cdot k$ is extra small, this presumably breaks down. At present, we do not know how this issue should be addressed.


\section{Scattering of self-dual dyons and black holes}
\label{sec:sd}

The preceding discussion deals with a potentially realistic effect. In this section, we comment instead on a purely academic problem, that of the scattering of self-dual dyons and black holes.\footnote{To our knowledge, the basic observations in this section were first made in the talk \cite{Monteiro:2023talk}.}

Self-duality, corresponding in our convention to
\[
\label{eq:sdEM}
e_i=ig_i
\]
for electromagnetic dyons, and to
\[
m_i=in_i
\]
for NUT-charged objects, requires a complexification. Self-dual fields must be complex in Lorentzian signature, namely (1,3). Natural settings for real self-dual fields are Euclidean and Kleinian (or split) signature, respectively (0,4) and (2,2). There is no notion of time and hence of real scattering in Euclidean signature, however. Here, we will remain agnostic and consider the complexified problem. One needs to specify a contour (e.g.~real momenta) to define an integral like the impulse \eqref{impulse}, but the feature we wish to highlight here applies already at the integrand level.

Let us start with electromagnetism. The U(1) electric-magnetic duality implies that observables transform covariantly under the duality. For instance, the impulse is invariant, while the waveform --- i.e.~$F_{\mu\nu}$ --- is covariant with degree 1. Interactions among two dyons, with straightforward extension to any number of dyons, depend on the duality-invariant charge combinations,
\[
e_1e_2 + g_1g_2,
\qquad e_1g_2 - e_2g_1.
\]
Clearly, both charge combinations vanish under the self-duality condition \eqref{eq:sdEM}. Therefore, the impulse vanishes to all perturbative orders. As a consequence, so does the waveform. There is no scattering. The same conclusion applies, of course, when both dyons are anti-self-dual.

An intuitive way to understand this result is the following. The self-duality condition means that the dyon couples only to one of the helicities of the photon. As seen in the 3-point amplitude \eqref{eq:3ptEM}, the complex charge is $e+i\eta g$. This implies that the expression \eqref{eq:A4sub} vanishes when both dyons are self-dual, because for each of the terms in the sum over helicities, $\eta\in\{+,-\}$, one of the two 3-point amplitudes vanishes. Diagramatically, looking at \eqref{eq:4ptdiag}, the would-be exchanged photon must have opposite helicities at each end, so cannot couple to one of the dyons.

Yet another way to understand this is by looking at the equation of motion for a dyon (a spinless one, for simplicity), which is of course consistent with the impulse. We have
\begin{equation}
m\,\frac{d u^\mu}{d\tau}
= (e F^{\mu\nu} + g\star\!F^{\mu\nu})u_\nu = \big( (e-ig)F^{\mu\nu}_\text{sd} + (e+ig)F^{\mu\nu}_\text{asd} \big)u_\nu,
\end{equation}
where $\star$ denotes the Hodge star, and (a)sd denote the (anti-)self-dual parts of $F^{\mu\nu}$. It is clear that a self-dual particle's trajectory is insensitive to the field sourced by another self-dual particle.

In gravity, more care is needed when taking the self-dual limit, because ${\mathcal M}_i=\sqrt{m_i^2+n_i^2}\rightarrow 0$. Since $p_i^\mu= {\mathcal M}_i u^\mu_i$, the momentum vanishes! One may approach this problem by using the impulse formula \eqref{eq:impgrav} to calculate $\Delta u^\mu_i$, not $\Delta p^\mu_i$. It is simpler, however, to look at the equation of motion for a probe self-dual particle (which we take here to be spinless for simplicity). This will be a nutty extension of the geodesic equation, which we have control over at leading perturbative order, because the gravitational U(1) duality holds. Such an extension of the geodesic equation was already considered in \cite{Emond:2020lwi}, where the extension of the geodesic deviation equation was also discussed; in fact, the former was obtained by integrating the latter. To leading perturbative order around flat space, the geodesic deviation obeys 
\begin{equation}
\frac{D^2 Y^\mu}{d\tau^2}
= \left( \frac{m}{{\mathcal M}}R^{\mu\nu}{}_{\rho\sigma} + \frac{n}{{\mathcal M}} \star R^{\mu\nu}{}_{\rho\sigma} \right) u_\nu u^\rho Y^\sigma = \left( \frac{m-in}{{\mathcal M}}R^{\mu\nu}_\text{sd}{}_{\rho\sigma} + \frac{m+in}{{\mathcal M}} R^{\mu\nu}_\text{asd}{}_{\rho\sigma} \right) u_\nu u^\rho Y^\sigma.
\end{equation}
Let us define the self-dual limit of the probe as $i n=m(1+\sigma)$ with $\sigma\rightarrow 0$. Hence, $m-in\propto\sigma$, while ${\mathcal M}\propto \sqrt\sigma$, so that $(m-in)/{\mathcal M}$ vanishes in the limit.
We conclude that the scattering of self-dual black holes is trivial, at least to leading order. Given that self-dual gravity is integrable, this is perhaps not surprising.\footnote{A side remark about amplitudes: as defined in \eqref{eq:3ptgrav}, the 3-point gravity amplitudes vanish for {\it both} helicities in the self-dual limit, due to their scaling with $(m+i\eta n){\mathcal M}$. This is a matter of convention, however. The amplitudes appear in observables as $\delta(2p\cdot q)\cl{A}_3^\text{grav} = \delta(2u\cdot q)\hat{\cl{A}}_3^\text{grav}$, where $\hat{\cl{A}}_3^\text{grav}:=\cl{A}_3^\text{grav}/{\mathcal M}$ vanishes only for one helicity.}

The self-dual Taub-NUT metric has a long history as a real solution in Euclidean signature \cite{Hawking:1976jb}. A recent development, motivated by celestial holography, is that it also has a role as a real solution in Kleinian signature, where it provides a notion of self-dual black hole \cite{Crawley:2021auj,Crawley:2023brz}. See also \cite{Adamo:2023fbj,Adamo:2025fqt,Guevara:2023wlr,Guevara:2024edh,Guevara:2025psg,Kim:2024dxo,Kim:2024mpy,Skvortsov:2025ohi,Kim:2025xka,Adamo:2026obu} for related recent work.
Now, in Euclidean signature, self-dual Taub-NUT centres can be superposed as in the Gibbons-Hawking multi-centred solution \cite{Gibbons:1978tef}. This solution has a Killing vector corresponding to `Euclidean time'. Considering its complexification, the centres may be thought of as being `at rest' with respect to each other. If the statement that `self-dual black holes' do not scatter is exact, then this suggests that they may be superposed even when they have relative velocities, which could correspond to a generalisation with no isometries of the complexified Gibbons-Hawking solution. However, the Gibbons-Hawking construction, as presently understood, relies on the existence of the `time' Killing vector. So, can such a generalised solution be constructed? Given the integrability of the equations of motion of self-dual gravity, this is a question that should be answerable.

We started this section by saying that the scattering of self-dual black holes is a purely academic problem. It is worth noting, however, that there have been attempts to describe the dynamics of a Lorentzian rotating (Kerr) black hole in terms of a pair of Taub-NUT instantons \cite{Kim:2024mpy,Kim:2025xka}. To our knowledge, truly dynamical aspects (requiring the Compton amplitude) in the Lorentzian regime have not yet been captured in this approach, though there is promising progress \cite{Kim:2026opo,Kim:2026yqo}.

\section{Conclusion}
\label{sec:conclusion}

We have investigated the memory effect resulting from the classical scattering of NUT-charged objects. This is the type of problem that would (at present) be hard to formulate in numerical relativity, so it is a natural target for scattering amplitudes methods. Nevertheless, it would be important to test these results using alternative approaches.

Let us summarise our discussion. The standard memory effect is captured by the soft behaviour of scattering amplitudes \cite{Strominger:2014pwa}, in particular by the Weinberg soft factor \cite{Weinberg:1964ew,Weinberg:1965nx}. We have studied here the generalisation of the soft factor to include NUT charges, and the consequences for the memory effect. The latter is encoded in the memory tensor $\mathcal{E}_{AB}(\hat x)$, which lives on the celestial sphere, such that the memory effect is the following displacement of distant freely falling test bodies, initially at fixed separation $\xi^B$ on the celestial sphere:
\[
\Delta \xi_A = \frac{\mathcal{E}_{AB}(\hat x)}{4\pi r}\,\xi^B.
\]
We will denote \,$
 E_{ABCD}\equiv \gamma_{AC}\gamma_{BD}+\gamma_{AD}\gamma_{BC}-\gamma_{AB}\gamma_{CD}
$,
and $\,O_{ABCD} \equiv \varepsilon_{AC}\gamma_{BD}+\varepsilon_{BD}\gamma_{AC}$, which will be associated below to the ``electric" and ``magnetic" components of the memory.

\begin{itemize}
\item The Weinberg soft factor for two-body scattering ($i=1,2$) is
\[
S_\text{Weinberg}^{(\eta)} = \frac{\kappa}{2}\sum_{i} \epsilon^{(\eta)}_{\mu}\epsilon^{(\eta)}_{\nu}\left(\frac{p_i'^{\mu} p_i'^{\nu}}{p_i'\cdot k} - \frac{p_i^\mu p_i^\nu}{p_i\cdot k}\right).
\]
This results in the memory tensor
\[
{\mathcal E}_{AB}= \frac{\kappa^2}{8}\sum_i E_{ABCD}D^C D^D\big((\ell\cdot \Delta p_i)\,\log\!\left(p_i\cdot \ell\right)\big) \,,
\]
where $D_A$ is a covariant derivative on the celestial sphere, and $\ell=(1,\hat x)$ identifies the direction of the observer measuring the memory. There is no magnetic component of the memory, which is consistent with expectations for conventional matter from traditional GR methods; see e.g.~\cite{Bieri:2018asm}.\footnote{An energy-momentum tensor that generates a magnetic component was devised in \cite{Satishchandran:2019pyc}.}

\item If at least one of the two bodies has a NUT charge, the naive modification of the Weinberg soft factor is
\[
S_\text{naive}^{(\eta)}= \frac{\kappa}{2}\sum_{i} e^{i\eta\theta_i}\,\epsilon^{(\eta)}_{\mu}\epsilon^{(\eta)}_{\nu}\left(\frac{p_i'^{\mu} p_i'^{\nu}}{p_i'\cdot k} - \frac{p_i^\mu p_i^\nu}{p_i\cdot k}\right), \quad \text{with} \quad e^{i\theta_i}=\frac{m_i+in_i}{\sqrt{m_i^2+n_i^2}}.
\]
Note that the momenta here are such that $p_i^\mu=\sqrt{m_i^2+n_i^2} \,u_i^\mu $. 
Except for this point, the expression above is analogous to the one that applies to dyons in electromagnetism. In gravity, however, the naive modification breaks gauge invariance generically. If we nonetheless proceed with the steps to determine the memory tensor, the result is a very natural extension of the standard case:
\[
{\mathcal E}_{AB}= \frac{\kappa^2}{8}\sum_i \Big[\cos\theta_i\, {E}_{ABCD} -\sin\theta_i\,{O}_{ABCD}  \Big] D^C D^D \big((\ell\cdot \Delta p_i)\,\log\!\left(p_i\cdot \ell\right)\big) \,,
\]
which now features a magnetic component. To obtain this result, we have made the gauge choice \eqref{eq:polsphere}, namely that the polarisation vectors are tangent to the celestial sphere.

\item Recent developments in scattering amplitudes motivate a gauge-invariant completion of the nutty soft factor:
\[
S_\text{grav}^{(\eta)} = \frac{\kappa}{2}\sum_i e^{i\eta\theta_i}\,\epsilon^{(\eta)}_{\mu}\epsilon^{(\eta)}_{\nu}\left(\frac{p_i'^{\mu} p_i'^{\nu}}{p_i'\cdot k} - \frac{p_i^\mu p_i^\nu}{p_i\cdot k}- \frac{q_i^\mu q_i^\nu}{q_i\cdot k}\right),
\]
where $q_i=p'_i-p_i$, and we have $q_1=-q_2$ as $k\rightarrow0$. If $\theta_i=0$, the extra terms cancel and we recover the Weinberg soft factor. The memory tensor is
\[
{\mathcal E}_{AB}= \frac{\kappa^2}{8}\sum_i \Big[\cos\theta_i\, {E}_{ABCD} -\sin\theta_i\,{O}_{ABCD}  \Big] D^C D^D \Phi_i \,,
\]
now with
\[
\Phi_i =
(\ell\cdot \Delta p_i)\,\log\!\left(\frac{p_i\cdot \ell}{|\ell_\perp|}\right) + \varepsilon_\perp(\ell_\perp,\Delta p_i)\arctan\!\left(\frac{\varepsilon_\perp(\ell_\perp,\Delta p_i)}{\ell\cdot \Delta p_i}\right) \,.
\]
Here, the subscript $\perp$ denotes a projection to the two-dimensional subspace orthogonal to $p_1$ and $p_2$; note that $\ell\cdot \Delta p_i=\ell_\perp\cdot \Delta p_i$, so the argument of the arctan is an angle. However, the memory tensor now exhibits a puzzling feature: it diverges as $|\ell_\perp|\rightarrow 0$, meaning in two antipodal directions; this is signalled in $\Phi_i$ by non-analyticity on the celestial sphere. It indicates a breakdown of the perturbation theory or of the soft approximation in these directions, which would be important to understand. In addition to this feature, our method is unable to determine a possible distributional piece supported at $\varepsilon_\perp(b,\ell_\perp)=0$, where $b$ is the impact parameter.
\end{itemize}

We emphasise that the puzzle between gauge invariance of the soft factor and regularity of the memory tensor is entirely absent from the story for electromagnetic dyons. Our paper was partly motivated by the realisation that there is a gauge-invariant completion of the nutty soft factor. However, it would be important to consider alternative ways in which the issue of gauge invariance can be addressed, e.g.~starting from position-space asymptotic methods.

While the discussion above is the focus of our paper, we have also briefly addressed an even more exotic problem, that of the scattering of self-dual black holes in complexified spacetime. Our observation that the scattering is trivial at leading perturbative order may feed into recent developments in celestial holography, where self-dual gravity has served as an interesting toy model.

There are various possible directions for future work, of which we highlight two (work in progress). The most obvious one is to consider subleading orders in the soft expansion, which correspond to the late-time waveform. This builds on extensive literature on (non-dyonic) classical soft theorems, e.g.~\cite{Laddha:2018rle,Laddha:2018myi,Sahoo:2018lxl,Saha:2019tub,Laddha:2019yaj,Sahoo:2020ryf,Sahoo:2021ctw,Manu:2020zxl,Sen:2024qzb,Alessio:2024onn,Choi:2024ygx,Karan:2025ndk,Akhtar:2025fil}. In fact, there is a recent derivation of the dyonic corrections to the all-order classical soft theorems in electromagnetism \cite{Duary:2025zqq}.\footnote{We thank the authors of \cite{Duary:2025zqq} for sharing their paper with us in advance of publication, after becoming aware of our independent result on the subleading classical soft theorem for dyons, presented in the talk \cite{Moynihan:2025talk}.} The gravity case is the natural next target.

The other obvious direction is the study of the bound problem, as opposed to the unbound / scattering problem of our paper. We expect that the small-frequency approximation provides useful information on binary dynamics in the presence of NUT charge, though the puzzle at $|\ell_\perp|\rightarrow0$ is a possible obstacle.

It is perhaps the case that this paper raises more questions than it answers. But if future gravitational wave observatories ever measure magnetic memory, those will become pressing questions.

\subsection*{Acknowledgements}

We are grateful to Francesco Alessio, Graham Brown, Carlo Heissenberg, Jung-Wook Kim, Lionel Mason and Donal O'Connell for discussions. We also thank Carlo Heissenberg and Jung-Wook Kim for comments on the manuscript. RM and GD acknowledge support from the Royal Society via a University Research Fellowship and an associated studentship grant, respectively. This work was also supported by the UK's Science and Technology Facilities Council (STFC) Consolidated Grants ST/T000686/1 and ST/X00063X/1 ``Amplitudes, Strings \& Duality".

\appendix

\section{Conventions and the Newman--Penrose formalism}\label{NPFormalism}

We work with metric signature $(+---)$ and introduce a null tetrad~\cite{Newman:1961qr}
\[
\{\ell^\mu,\tilde{\ell}^\mu,m^\mu,\bar{m}^\mu\},
\]
normalised by
\[
\ell\cdot\tilde{\ell}=1,\qquad m\cdot\bar{m}=-1,
\]
with all other inner products vanishing. With the conventions
\[
\label{eq:symcon}
A_{(\mu}B_{\nu)}\equiv A_\mu B_\nu + A_\nu B_\mu,\qquad
A_{[\mu}B_{\nu]}\equiv A_\mu B_\nu - A_\nu B_\mu,
\]
the metric is
\[
\eta_{\mu\nu}=\ell_\mu \tilde{\ell}_\nu+\tilde{\ell}_\mu \ell_\nu
          -m_\mu \bar{m}_\nu-\bar{m}_\mu m_\nu
      \;=\;\ell_{(\mu}\tilde{\ell}_{\nu)}-m_{(\mu}\bar{m}_{\nu)} .
\]
Geometrically, $\ell^\mu$ points outward along the radiation direction, $\tilde{\ell}^\mu$ inward, and
$m^\mu,\bar m^\mu$ span the transverse polarization plane.

Let $x^A=(z,\bar z)$ be complex coordinates on the celestial sphere, and take $\ell^\mu=\ell^\mu(z,\bar z)$ to be the corresponding null direction field, which may be parametrised as
\[
\ell^\mu = \frac1{1+z\bar z}\big(1+z\bar z,z+\bar z,i(\bar z- z),1-z\bar z\big).
\]
The tangent vectors to $S^2$,
\[
e_A^{\mu}\equiv \partial_A \ell^\mu,\qquad
(e_z^\mu,e_{\bar z}^\mu)=(m^\mu,\bar m^\mu),
\]
induce the (positive-definite) sphere metric via
\[
e_A\cdot e_B \equiv \eta_{\mu\nu}e_A^\mu e_B^\nu = -\gamma_{AB}.
\]
The inverse zweibein is
\[
e^A{}_\mu \equiv -\gamma^{AB} \eta_{\mu\nu} e_B^\nu,
\]
so that $e^A{}_\mu e_B^\mu=\delta^A{}_B$ and, equivalently, $e^{A\mu}=-\gamma^{AB}e_B^\mu$.

We introduce a complex dyad $(m^A,\bar m^A)$ on $S^2$ obeying
\[
m_A m^A=0,\qquad m_A\bar m^A=1,
\]
so that
\[
m^\mu=e_A^\mu m^A,\qquad \bar m^\mu=e_A^\mu \bar m^A,
\]
and
\begin{equation}\label{spheremetricdyads}
\gamma_{AB}=m_{(A}\bar m_{B)},\qquad
\varepsilon_{AB}= i\, m_{[A}\bar m_{B]} .
\end{equation}
With the tetrad and dyad in hand, helicity eigenstate polarization vectors may be obtained as
\[
\label{eq:polsphere}
\epsilon_\mu^{(+)}=m_\mu,\qquad \epsilon_\mu^{(-)}=\bar m_\mu,
\qquad\text{so that}\qquad
\epsilon_\mu^{(\eta)} = m_A^{(\eta)}\,e^A{}_\mu .
\]

\section{Useful integral}
\label{app:locus}
Using the identities
\[
\frac{1}{x+i0}=-i\int_0^\infty ds\,e^{isx},
\qquad
\int d^2q_\perp\,e^{ik_\perp\cdot q_\perp}=(2\pi)^2\delta^{(2)}(k_\perp),
\]
we can write
\[
\int d^2q_\perp\,\frac{e^{ib_\perp\cdot q_\perp}}{\ell_\perp\cdot q_\perp+i0}
=-i(2\pi)^2\int_0^\infty ds\,\delta^{(2)}(b_\perp+s\ell_\perp)
=-4\pi^2 i\,\theta(-b_\perp\!\cdot\!\ell_\perp)\,
\delta\!\bigl(\varepsilon_\perp(b,\ell)\bigr),
\]
where we recall that $b=b_\perp$.

\bibliographystyle{JHEP}
\bibliography{mainbib} 
\end{document}